# Graphene, neutrino mass and oscillation

Zhong-Yue Wang

*Research Center, SBSC, No. 580, Songhuang Road*

*Shanghai 201703, China*

*E-mail: zhongyuewang(at)yahoo(dot)com(dot)cn*

**Abstract:** In 2007 a sonic Michelson-Morley experiment was carried out and the result is also null[1] consistent with the hypothesis to establish a system similar to special relativity provided the light speed $c$ is replaced by the constant speed $c_s$ of mechanical waves[2]. Thereby we interpreted why the energy-momentum relation $E = pc_s \propto p$ of a phonon seems like a photon, the extreme relativistic particle. Furthermore, we thought other constant velocities can play the role of $c$ as well. Recently, electrons in graphene were discovered to behave as the coefficient is a Fermi velocity $V_F$[3]. Finally, a conjecture for neutrinos to avoid the contradiction among two-component theory, negative rest mass-square [4] and oscillation is presented.

## 1. Introduction

Two main theories to the energy-momentum tensor of an electromagnetic field in media were proposed by Minkowski (1908) and Abraham (1909) [5]. Minkowski's momentum $En/c = hfn/c > hf/c$ accords with de Broglie formula $p = h/\lambda$ of quantum mechanics. But Abraham's expression $E/cn = hf/cn < hf/c$ can be deduced from the theory of relativity( $p = mV, m = E/c^2, V = c/n$ ). How to resolve this puzzle?

Actually, mechanical equations of relativity

$$E = mc^2 = \frac{m_0 c^2}{\sqrt{1 - V^2/c^2}}$$

$$p = mV = \frac{m_0 V}{\sqrt{1 - V^2/c^2}}$$

$$\frac{E}{p} = \frac{c^2}{V}$$

$$E^2 = p^2 c^2 + m_0^2 c^4$$

are merely tenable to free particles. They cannot be applied to describe photons in media due to interactions between the field and matter.

## 2. Solution

Permittivity $\varepsilon$ and permeability $\mu$ of media are different to $\varepsilon_0$ and $\mu_0$ in vacuum which has contained the mentioned interaction. Therefore, we try to introduce $u = 1/\sqrt{\varepsilon\mu}$ to take the place of $c = 1/\sqrt{\varepsilon_0 \mu_0}$

$$E = mu^2 = \frac{m_0 u^2}{\sqrt{1 - V^2/u^2}}$$

$$p = mV = \frac{m_0 V}{\sqrt{1 - V^2/u^2}}$$

$$\frac{E}{p} = \frac{u^2}{V}$$

$$E^2 = p^2 u^2 + m_0^2 u^4$$

to study particles in a medium as if they are free because the interaction is equivalent to the change of $1/\sqrt{\varepsilon_0 \mu_0}$ to $u = 1/\sqrt{\varepsilon\mu}$. The correctness can be judged by comparison with other theories and experiments. For instance, the energy-mass and energy-momentum relation of a photon within media is

$$\frac{E}{m} = u^2 = \frac{1}{\varepsilon\mu}$$

$$\frac{E}{p} = \frac{u^2}{V} = \frac{u^2}{u} = u = \frac{1}{\sqrt{\varepsilon\mu}}$$

whose velocity $V$ is the energy flow velocity $1/\sqrt{\varepsilon\mu}$. They are in agreement with following outcomes of electrodynamics

$$\frac{w}{M} = \frac{w\mathbf{V}}{M\mathbf{V}} = \frac{\mathbf{S}}{\mathbf{g}} = \frac{\mathbf{E}\times\mathbf{B}/\mu}{\varepsilon\mathbf{E}\times\mathbf{B}} = \frac{1}{\varepsilon\mu} = u^2$$

$$\frac{w}{g} = \frac{\frac{1}{2}(\varepsilon E^2 + \frac{1}{\mu}B^2)}{\varepsilon EB} = \frac{\varepsilon E^2}{\varepsilon EB} = \frac{1}{\sqrt{\varepsilon\mu}}$$

( $w$_energy density, $M$_mass density, $\mathbf{V}$_velocity, $\mathbf{S}$_Poynting's vector, $\mathbf{g}$_momentum density)

In vacuum, $u = c = 1/\sqrt{\varepsilon_0\mu_0}$ and they reduce to known equations of relativity. Minkowski's formula $p = nE/c$ is correct.

## 3. Cherenkov effect

These equivalent mechanical equations are still valid to massive particles. Consider the radiation of an electron moving in the medium.

Momentum conservation: $\mathbf{p}_1 - \hbar\mathbf{k} = \mathbf{p}_2$

$\mathbf{p}_1$ is the initial momentum of the particle and $\hbar\mathbf{k}$ is the momentum of the emitted photon. Thus,

$$p_1^2 - 2p_1\hbar k \cos\theta + \hbar^2 k^2 = p_2^2$$

$$\cos\theta = \frac{p_1^2 + \hbar^2 k^2 - p_2^2}{2p_1\hbar k}$$

Meanwhile, energy conservation is
$$E_1 - \hbar\omega = E_2$$

If the energy is $E = \sqrt{m_0^2 u^4 + p^2 u^2}$,
$$m_0^2 u^4 + p_1^2 u^2 + (\hbar\omega)^2 - 2\hbar\omega E_1 = m_0^2 u^4 + p_2^2 u^2$$

$$p_1^2 + (\hbar\omega/u)^2 - 2\hbar\omega E_1/u^2 = p_2^2$$

Substituting it into the expression of the angle,

$$\cos\theta = \frac{\hbar^2 k^2 - (\hbar\omega/u)^2 + 2\hbar\omega E_1/u^2}{2p_1\hbar k}$$

Owing to $u = \omega/k = c/n$ and $E_1/p_1 = u^2/V_1$,

$$\cos\theta = \frac{2\hbar\omega E_1/u^2}{2p_1\hbar k} = \frac{\omega/k}{V_1} = \frac{c/n}{V_1}$$

In the case of $V_1 < c/n$, clearly, $\cos\theta > 1$ and the radiation is forbidden. The Cherenkov effect occurs when the charged particle is faster than the phase velocity $\omega/k$.

## 4. Transition radiation

According to the equivalent theory, the energy to a charged particle in the medium $n_1$ and $n_2$ is

$$E_1 = \sqrt{m_0^2 u_1^4 + p_1^2 u_1^2} \qquad (u_1 = c/n_1)$$

$$E_2 = \sqrt{m_0^2 u_2^4 + p_2^2 u_2^2} \qquad (u_2 = c/n_2)$$

There should exist a radiation to keep energy-momentum conservation while the particle entering $n_2$ from $n_1$. It is just the transition radiation pointed out by Frank and Ginzburg in 1940s. Here,

$$E_1 = E_2 + \hbar\omega \qquad (\hbar k = \frac{\hbar\omega}{u_2} = \frac{n_2 \hbar\omega}{c})$$

$$m_0^2 u_1^4 + p_1^2 u_1^2 + (\hbar\omega)^2 - 2\hbar\omega E_1 = m_0^2 u_2^4 + p_2^2 u_2^2$$

$$m_0^2 \frac{u_1^4}{u_2^2} + p_1^2 \frac{u_1^2}{u_2^2} + (\hbar k)^2 - \frac{2\hbar\omega E_1}{u_2^2} = m_0^2 u_2^2 + p_2^2$$

$$p_1^2 + (\hbar k)^2 - p_2^2 = m_0^2(u_2^2 - \frac{u_1^4}{u_2^2}) + p_1^2(1 - \frac{u_1^2}{u_2^2}) + \frac{2\hbar\omega E_1}{u_2^2}$$

$$\cos\theta = \frac{m_0^2 c^2(\frac{1}{n_2^2} - \frac{n_2^2}{n_1^4}) + p_1^2(1 - \frac{n_2^2}{n_1^2})}{2p_1\hbar\omega n_2/c} + \frac{c/n_2}{V_1}\frac{n_2^2}{n_1^2}$$

which is the Cherenkov angle $\cos\theta = \dfrac{c/n}{V_1}$ on the condition of $n_1 = n_2 = n$.

## 5. Graphene and superconductors

The Fermi velocity of condensed matter physics can take the place of $c$ too [3]. So, the energy gap is treated as $E_g = 2m_0c^2 \to 2m_0V_F^2$ in view of Dirac's theory to antiparticles and the equivalent equation to electrons in an uniform electric field $F$ should be

$$\left\{\alpha\, V_F\mathbf{p} + \beta\dfrac{E_g}{2} - eFz - E\right\}\psi = 0$$

The exact solution of

$$\left\{\alpha\, c\mathbf{p} + \beta m_0c^2 - eFz - E\right\}\psi = 0$$

had been given [6] and $c \longleftrightarrow V_F$, $m_0c^2 \longleftrightarrow \dfrac{E_g}{2}$. The tunneling probability[7]

$$P = \dfrac{e^2 F^2}{4\pi^3 \hbar^2 c}\sum_{n=1}^{\infty}\dfrac{1}{n^2}\exp(-\dfrac{n\pi\, m_0^2 c^3}{e\hbar F})$$

is now [8]

$$P \propto \sum_{n=1}^{\infty}\dfrac{1}{n^2}\exp(-\dfrac{n\pi E_g^2}{4e\hbar F V_F})$$

On the other hand, the Unruh effect

$$k_B T = \dfrac{\hbar}{2\pi c}a \to \dfrac{\hbar}{2\pi V_F}a$$

of general relativity ( $k_B$ _Boltzman constant; $T$_ temperature; $a$_acceleration) is helpful to estimate the coherent length $\xi$ of superconducting carriers. The energy gap $E_g = 2\Delta$ equals twice of the rest energy of a quasiparticle $m_0c^2 \to m_0V_F^2$. Consequently, the force between two particles is

$$m_0 a_c = \dfrac{\Delta}{V_F^2}\dfrac{2\pi V_F k_B T_c}{\hbar} = \dfrac{2\pi\Delta k_B T_c}{\hbar V_F}$$

The work is $m_0 a_c \xi = E_g$ and

$$\xi = \dfrac{E_g}{m_0 a_c} = \dfrac{2\Delta}{m_0 a_c} = \dfrac{\hbar V_F}{\pi k T_c}$$

has the same order of magnitude of

$$\xi = \dfrac{\hbar V_F}{\Delta} \xrightarrow{2\Delta = 3.53 kT_c} \dfrac{\hbar V_F}{1.76 kT_c}$$

in the BSC theory.

The frequency of the a.c. Joesphson effect is

$$\dfrac{\partial}{\partial t}(\varphi_1 - \varphi_2) = \dfrac{2eU}{\hbar} = \dfrac{2\int e(F_1 - F_2)dl}{\hbar}$$

When the voltage is zero and a heat flow is across the junction,

$$\dfrac{\partial}{\partial t}(\varphi_1 - \varphi_2) = \dfrac{2\int m_0(a_1 - a_2)dl}{\hbar} = \dfrac{4\pi k\Delta}{\hbar^2 V_F}\int(T_1 - T_2)dl$$

$\Delta$ is a variation related to the temperature and a more precise formula to $T_1 \neq T_2$ should be

$$\dfrac{\partial}{\partial t}(\varphi_1 - \varphi_2) = \dfrac{2\int (m_1 a_1 - m_2 a_2)dl}{\hbar} = \dfrac{4\pi k}{\hbar^2 V_F}\int(\Delta_1 T_1 - \Delta_2 T_2)dl$$

Such a frequency is in the radio region[9]. This cannot be interpreted to be the thermoelectric effect because an extra phase rather than current is induced. The temperature-phase phenomenon can be regarded as the analogue of the converse Unruh effect in condensed matter physics. It should be dependent on temperature gradient and not the quantity of heat.

## 6. Optical experiment to test the hypothesis

The addition rule of collinear velocities is

$$v_{sum} = \dfrac{v_1 + v_2}{1 + \dfrac{v_1 v_2}{c^2}}$$

Hence, in Fizeau's experiment (1850s),

$$\frac{\frac{c}{n} \pm V_w}{1 \pm \frac{c}{n} \frac{V_w}{c^2}} \approx (\frac{c}{n} \pm V_w)(1 \mp \frac{V_w}{cn}) \approx \frac{c}{n} \pm (1-\frac{1}{n^2})V_w$$

where $v_1 = \frac{c}{n}$ is the light speed in still water,

$v_2 = \pm V_w$ is the velocity of water and $1-\frac{1}{n^2}=k$

is the Fresnel drag coefficient. Shifts of interference fringes are

$$\Delta N = \frac{\Delta t}{T} = f(\frac{2L}{\frac{c}{n}-kV_w} - \frac{2L}{\frac{c}{n}+kV_w}) \approx \frac{4Ln^2 fkV_w}{c^2}$$

Postulate to do the experiment in a transparent medium whose refractive index is $n'=c/u$. The transformation corresponding to those equivalent mechanical equations is

$$\begin{pmatrix} x' \\ y' \\ z' \\ t' \end{pmatrix} = \begin{pmatrix} \frac{1}{\sqrt{1-\frac{V^2}{u^2}}} & 0 & 0 & \frac{-V}{\sqrt{1-\frac{V^2}{u^2}}} \\ 0 & 1 & 0 & 0 \\ 0 & 0 & 1 & 0 \\ \frac{-V/u^2}{\sqrt{1-\frac{V^2}{u^2}}} & 0 & 0 & \frac{1}{\sqrt{1-\frac{V^2}{u^2}}} \end{pmatrix} \begin{pmatrix} x \\ y \\ z \\ t \end{pmatrix}$$

and reduced to the Lorentz transformation in vacuum. Herewith, the addition rule of velocities will be

$$\frac{v_1+v_2}{1+\frac{v_1 v_2}{u^2}} \quad (u=\frac{c}{n'})$$

In this experiment,

$$v_{sum} = \frac{\frac{c}{n} \pm V_w}{1 \pm \frac{\frac{c}{n} V_w}{n'^2}} \approx (\frac{c}{n} \pm V_w)(1 \mp \frac{V_w}{cn}n'^2) \approx \frac{c}{n} \pm (1-\frac{n'^2}{n^2})V_w$$

The drag coefficient is now $k=1-\frac{n'^2}{n^2}$. In vacuum or air, $n' \to 1$ and the coefficient is Fresnel's $k \approx 1-\frac{1}{n^2}$.

## 7. Test in a gravitational field

The metric of the extended Lorentz transformation is

$$ds^2 = dr^2 + r^2 d\theta^2 + r^2 \sin^2\theta \, d\varphi^2 - u^2 dt^2$$

Obviously, in a gravitational field

$$ds^2 = \frac{dr^2}{1-\frac{2GM}{u^2 r}} + r^2 d\theta^2 + r^2 \sin^2\theta \, d\varphi^2 - (1-\frac{2GM}{u^2 r})u^2 dt^2$$

$$\frac{f_1}{f_2} = \frac{\sqrt{1-\frac{2GM}{u^2 r_2}}}{\sqrt{1-\frac{2GM}{u^2 r_1}}} \approx \frac{1-\frac{GM}{u^2 r_2}}{1-\frac{GM}{u^2 r_1}}$$

On the surface of the earth, $\Delta f/f \approx gH/u^2$ is in accordance with the result of Newton's theory.

$$\Delta E = mgH$$
$$E=hf, \quad \Delta E = h\Delta f, \quad m=E/u^2 = E\varepsilon\mu$$
$$h\Delta f = \frac{hf}{u^2} gH$$
$$\frac{\Delta f}{f} = \frac{gH}{u^2}$$

The shift depends upon the product $\varepsilon\mu$ instead of the phase velocity, group velocity, etc. The value of a γ-photon is approximate to $1/c^2$ in the well-known experiment of Pound, et al [10] and we suggest to measure wherein $1/\varepsilon\mu$ is low enough such as ultra-high dielectric constant media. As to a left-handed-material(LHM), the shift is reversed.

Likewise, it is interesting to test the gravitational shift of mechanical waves experimentally[2]. The mass-energy equation of a phonon is $E = mc_s^2$ [2] and

$$hf_1 - \frac{GM}{r_1}\frac{hf_1}{c_s^2} = hf_2 - \frac{GM}{r_2}\frac{hf_2}{c_s^2}$$

Generally speaking, $\frac{GM}{c_s^2 r} \gg 1$ so that

$$-\frac{GM}{r_1}\frac{hf_1}{c_s^2} = -\frac{GM}{r_2}\frac{hf_2}{c_s^2}$$

$$\frac{f_1}{f_2} = \frac{r_1}{r_2}$$

$$\frac{\Delta f}{f} \approx -\frac{H}{R} \quad (R = 6371\,km\text{\_radius of earth})$$

According to "general sonic relativity",

$$\frac{f_1}{f_2} = \frac{\sqrt{1 - \frac{2GM}{c_s^2 r_2}}}{\sqrt{1 - \frac{2GM}{c_s^2 r_1}}} = \frac{\sqrt{\frac{2GM}{c_s^2 r_2} - 1}}{\sqrt{\frac{2GM}{c_s^2 r_1} - 1}} \approx \sqrt{\frac{r_1}{r_2}}$$

$$\frac{\Delta f}{f} \approx -\frac{H}{2R}$$

The difference between the Newton theory is detectable.

## 8. Neutrinos

Assume the neutrino has a constant speed $c_v$. It is possible but unnecessary to be the light speed $c$. Thereafter, we can establish a mechanical system similar to special relativity and $c$ is replaced by $c_v$ whose formula should be

$$E = mc_v^2 = \frac{m_0 c_v^2}{\sqrt{1 - V^2/c_v^2}}$$

$$p = mV = \frac{m_0 V}{\sqrt{1 - V^2/c_v^2}}$$

$$\frac{E}{p} = \frac{c_v^2}{V}$$

$$E^2 = p^2 c_v^2 + m_0^2 c_v^4$$

The rest mass of a neutrino is zero and energy-momentum relation is $E = pc_v$ on account of $V = c_v$, and thus parity is non-conservational. Whereas, people are used to choose $E^2 = p^2 c^2 + m_0^2 c^4$ to fit experimental data

$$E^2 = \frac{E^2}{c_v^2}c^2 + m_0^2 c^4$$

that $m_0^2$ must be negative[4] if $c_v < c$. Maybe three flavours have different velocities, e.g. $c_{v\tau} < c_{v\mu} < c_{ve} < c$. In brief,

$c_{ve}$ is constant, $m_{0(ve)} = 0$, Parity violation;

$c_{v\mu}$ is constant, $m_{0(v\mu)} = 0$, Parity violation;

$c_{v\tau}$ is constant, $m_{0(v\tau)} = 0$, Parity violation;

$c_{v\tau} \neq c_{v\mu} \neq c_{ve}$ behaves as oscillation.

$c_v \approx c$ in light of the rest mass of $v_e$ from experiments is proximal to zero.

**References:**

[1]. Liu,W., Su,B., Xi,D., *et al.* Michelson-Morley Experiment in Sound Interference of Compressible Fluid, ***Mechanical Science and Technology for Aerospace Engineering***, volume26, No.,9, 1144-1146 (in Chinese)

[2]. Wang, Z., On the dynamics of mechanical waves and phonons, ***Journal of Pure and Applied Ultrasonics.*** **22(**4), 106-110(2000)

[3]. Novoselov, K.S., Gein, A.K., Morozov,S.V., *et al.* Two-dimensional gas of massless Dirac fermions in graphene *Nature* 438, 197–200(2005)

[4]. Review of Particles Physics, *Phys. Rev. D* 54,280(1996)

[5]. Bowyer,P., The momentum of light in media: the Abraham-Minkowski controversy (2005),
   http://peter.mapledesign.co.uk/writings/physics/2005_dissertation_The_Abraham-Minkowski_Controversy.pdf

[6]. Ley, K., Wang, R., Ren,S., *et al.* Charged spin-1/2 particles in uniform electrical field, *Journal of University of Science and Technology of China*, vol.**13**, No.2,167–181(1983)

[7]. Itzykson, C., Zuber, J-B., *Quantum Field Theory*, McGraw-Hill(1980), Chap.4-3-3

[8]. Jena, D., Fang, T., Zhang,Q., *et al.* Zener Tunneling in semiconducting nanotube and graphene nanoribbon p−n junctions, *Appl. Phys. Lett.* **93**, 112106(2008)

[9]. Panaitov,G.I., Ryazanov, V.V., Ustinov,A.V., *et al.*, Thermoelectric a.c.Josephson effect in SNS junctions, *Phys. Lett. A* **100**, 301-303(1984)

[10]. Pound, R.V., Rebka, G.A., Apparent weight of photons, *Phys Rev.Lett.* **4**, 337-341(1960)